\def\qquad{\hspace{30pt}}
\def\0{\newline}
\def\1{\newline}
\def\2{\par\noindent\newline}

\def\loq{,\kern-0.080em,\kern+0.05em}

\def\bfloq{{\bf,\kern-0.06em,\kern+0.05em}}

\def\footloq{,\kern-0.07em,\kern+0.03em}

\def\quabla{{\raise.7ex\hbox{\boxed{{}}}}}

\documentclass[german,12pt]{article}
 \usepackage[english]{babel}
 \usepackage{latexsym}
 \usepackage{graphicx}
 \usepackage{ifthen}
\title{}
\date{}
\begin{document}

%
%
\begin{center}
\                                                                         \\[0.00cm]
{\LARGE\bf On The Barometric Formulas\\
           And Their Derivation From\\
           Hydrodynamics and Thermodynamics\\[1.0cm]
}
{\large\bf Version 2.0 (March 9, 2010)}\\[2.0cm]
%
%
{\Large\sc Gerhard Gerlich}                                               \\[0.30cm]
{\large\rm Institut f\"ur Mathematische Physik}                           \\[0.30cm]
{\large\rm Technische Universit\"at Carolo-Wilhelmina zu Braunschweig}    \\[0.30cm]
{\large\rm Mendelssohnstra\ss e 3}                                        \\[0.30cm]
{\large\rm D-38106 Braunschweig}                                          \\[0.30cm]
{\large\rm Federal Republic of Germany}                                   \\[0.30cm]
{\large\rm g.gerlich@tu-bs.de}                                            \\[1.50cm]
%
%
{\Large\sc Ralf D. Tscheuschner}                                          \\[0.30cm]
{\large\rm Postfach 60\,27\,62}                                           \\[0.30cm]
{\large\rm D-22237 Hamburg}                                               \\[0.30cm]
{\large\rm Federal Republic of Germany}                                   \\[0.30cm]
{\large\rm ralfd@na-net.ornl.gov}                                         \\[0.30cm]
%
%
\end{center}
%
%
\pagebreak
%
%
\ \\[4cm]
\begin{center}
{\large\bf Abstract}\\
\end{center}
We derive the approximate pressure profiles, density profiles, and
temperature profiles of an atmosphere, also called {\it barometric
formulas\/}. Our variant of a derivation goes beyond the common
standard exercise of a thermodynamics lecture, where commonly the
discussion of the underlying physical assumptions is missed. We
depart from the Navier-Stokes equation and explicitly point our
attention on the physical assumptions disregarded elsewhere. We show
that the usual assumptions can be relaxed leading to generalized
formulas that hold even in the case of horizontal winds. This
fundamental physics has some relevance to the current discussions on
the climate debate.
\pagebreak
%
%
\tableofcontents
\newpage
\pagestyle{myheadings}
\markboth{%
{\sl Gerlich and Tscheuschner, On The Barometric Formulas $\dots$ }}%
{{\sl Gerlich and Tscheuschner, On The Barometric Formulas $\dots$}}
%
%
\section{Introduction}
In the following, we derive approximate temperature profiles of an
atmosphere, also called adiabatic lapse rates or better {\it
barometric formulas\/}. Our variant of a derivation goes beyond the
common standard exercise of a thermodynamics lecture, where commonly
the discussion of the underlying physical assumptions is missed. We
depart from the Navier-Stokes equation and explicitly point our
attention on the physical assumptions disregarded elsewhere. By the
way, this derivation is a good example on how to apply the
magnetohydrodynamic equations regarded as redundant by some of our
critics. Furthermore, it explicitly shows that in physics an
application of formulas is valid {\it only in a finite space-time
region\/}. In addition, we show that the usual assumptions can be
relaxed leading to generalized formulas that hold even in the case
of horizontal winds.

A brief historical review of the barometric formula is given in
Ref.\ \cite{Berberan1996}. The reader is also referred to the
textbook by Riegel and Bridger on \lq\lq Fundamentals of Atmospheric
Dynamics and Thermodynamics\rq\rq\ \cite{Riegel1992}.
%
%
\section{On the derivation of the barometric formulas}
\subsection{Input from hydrodynamics}
As described in our falsification paper \cite{GT2007,GT2009} the
core of a climate model must be a set of equations describing the
equations of fluid flow, namely the generalized Navier-Stokes
equations. They describe the conservation of momentum and read
\begin{equation}
\frac{\partial}{\partial t} ( \varrho\,\textbf{v} )
+ \mbox{\boldmath$\nabla$} \cdot (
\varrho\,\textbf{v}\otimes\textbf{v} )
=
- \mbox{\boldmath$\nabla$} p
- \varrho \, \mbox{\boldmath$\nabla$} \Phi
+ \varrho_e \textbf{E}
+ \textbf{j} \times \textbf{B}
+ \mbox{\boldmath$\nabla$} \cdot \textbf{R}
+ \textbf{F}_{\mbox{\scriptsize\rm ext}}
\label{Eq:NSE}
\end{equation}
where $\textbf{v}$ is the velocity vector field, $p$ the pressure
field, $\Phi$ the gravitational potential, $\textbf{R}$ the friction
tensor, and $\textbf{F}_{\mbox{\scriptsize\rm ext}}$ are the
external force densities, which could describe the Coriolis and
centrifugal accelerations. Neglecting the friction term and the
electromagnetic fields we obtain the Euler equations.
\subsubsection*{Assumption 1}
\begin{itemize}
\item
We neglect the electromagnetic field terms.
\end{itemize}
We get the more common version of the Navier-Stokes equations
\begin{equation}
\frac{\partial}{\partial t} ( \varrho\,\textbf{v} )
+ \mbox{\boldmath$\nabla$} \cdot (
\varrho\,\textbf{v}\otimes\textbf{v} )
=
- \mbox{\boldmath$\nabla$} p
- \varrho \, \mbox{\boldmath$\nabla$} \Phi
+ \mbox{\boldmath$\nabla$} \cdot \textbf{R}
+ \textbf{F}_{\mbox{\scriptsize\rm ext}}
\end{equation}
The left hand side of this equation may be rewritten according to
\begin{equation}
\frac{ \partial }{ \partial t } ( \varrho\,\textbf{v} )
+ \mbox{\boldmath$\nabla$} \cdot (
\varrho\,\textbf{v}\otimes\textbf{v} )
=
\frac{ \partial\varrho }{ \partial t } \textbf{v} +
\varrho \cdot \frac{ \partial\textbf{v} }{ \partial t } +
\mbox{\boldmath$\nabla$} \cdot ( \varrho\textbf{v} ) \, \textbf{v} +
\varrho \textbf{v} \cdot \mbox{\boldmath$\nabla$} \, \textbf{v}
\end{equation}
With the continuity equation for the mass density
$\partial\varrho/\partial t + \mbox{\boldmath$\nabla$} \cdot
(\varrho\textbf{v}) = 0$ this term simplifies to
\begin{equation}
\varrho \cdot \frac{ \partial\textbf{v} }{ \partial t } +
\varrho \textbf{v} \cdot \mbox{\boldmath$\nabla$} \, \textbf{v}
\end{equation}
Thus we obtain the well-know form of the Navier-Stokes equations,
or, preferring the singular form, the Navier-Stokes equation
\begin{equation}
\varrho \cdot \frac{ \partial\textbf{v} }{ \partial t } +
\varrho \textbf{v} \cdot \mbox{\boldmath$\nabla$} \, \textbf{v}
=
- \mbox{\boldmath$\nabla$} p
- \varrho \, \mbox{\boldmath$\nabla$} \Phi
+ \mbox{\boldmath$\nabla$} \cdot \textbf{R}
+ \textbf{F}_{\mbox{\scriptsize\rm ext}}
\end{equation}
where the term $- \varrho \, \mbox{\boldmath$\nabla$} \Phi$ is
gravity. If we neglect the viscosity term, we are left with the
Euler equation
\begin{equation}
\varrho \cdot \frac{ \partial\textbf{v} }{ \partial t } +
\varrho \textbf{v} \cdot \mbox{\boldmath$\nabla$} \, \textbf{v}
=
- \mbox{\boldmath$\nabla$} p
- \varrho \, \mbox{\boldmath$\nabla$} \Phi
+ \textbf{F}_{\mbox{\scriptsize\rm ext}}
\end{equation}
\subsubsection*{Assumption 2}
\begin{itemize}
\item
We assume that $\textbf{v}(\textbf{r},t)$ is independent of
$\textbf{r}$. In sharp distinction to the standard derivation of the
barometric formulas, we relax the usual condition
$\textbf{v}(\textbf{r},t)\equiv 0$ in order to allow non-vanishing
velocity fields $\textbf{v}(\textbf{r},t)$, which are independent of
$\textbf{r}$.
\end{itemize}
Consequently, the viscosity tensor $\textbf{R}$ and the non-linear
term $\varrho\textbf{v}\cdot\mbox{\boldmath$\nabla$}\textbf{v}$ are
zero, such that
\begin{equation}
\varrho \cdot \frac{ \partial\textbf{v} }{ \partial t }
=
- \mbox{\boldmath$\nabla$} p
- \varrho \, \mbox{\boldmath$\nabla$} \Phi
+ \textbf{F}_{\mbox{\scriptsize\rm ext}}
\label{Eq:EE}
\end{equation}
{\it Remark:\/} If one writes
\begin{equation}
\varrho\,\textbf{v} \cdot \mbox{\boldmath$\nabla$} \textbf{v}
=
- \varrho\,\textbf{v} \times
( \mbox{\boldmath$\nabla$} \times \textbf{v} )
+
\varrho\,\mbox{\boldmath$\nabla$}
\left( \frac{1}{2} |\textbf{v}|^2\right)
\end{equation}
one could weaken this assumption to potential velocity fields. With
these formulas one can derive the Bernoulli equation.

In case of a rotating atmosphere of the Earth the last term
$\textbf{F}_{\mbox{\scriptsize\rm ext}}$ of the right hand side of
Eq.~\ref{Eq:EE} describes the centrifugal acceleration and the
Coriolis acceleration. The latter vanishes for a identically
vanishing velocity field.
\subsubsection*{Assumption 3}
\begin{itemize}
\item
We set $\textbf{F}_{\mbox{\scriptsize\rm ext}}$ to zero.
\end{itemize}
We now have two fields, namely $-\mbox{\boldmath$\nabla$}p$ and $-
\varrho \, \mbox{\boldmath$\nabla$} \Phi$, which will accelerate the
volume elements, if they are different fields:
\begin{equation}
\varrho \cdot \frac{ \partial\textbf{v} }{ \partial t }
=
- \mbox{\boldmath$\nabla$} p
- \varrho \, \mbox{\boldmath$\nabla$} \Phi
\label{Eq:Ass3}
\end{equation}
Let us follow the common notation and write for the gravitational
field
\begin{equation}
\textbf{g} = - \mbox{\boldmath$\nabla$} \Phi
\end{equation}
\subsubsection*{Assumption 4}
\begin{itemize}
\item
We assume that, as usual, acceleration due to gravity is vertical,
i.e. we set
\begin{equation}
\textbf{g} = - g\, \textbf{e}_z
\end{equation}
where $\textbf{e}_z$ is the unit vector in $z$-direction and $g$ is
constant in space and time. This is the flat earth hypothesis.
Furthermore, we neglect the variation of the gravitation field
induced by the gravitation fields of the Sun, the Moon, and the
planets.
\end{itemize}
\subsubsection*{Assumption 5}
\begin{itemize}
\item
We assume that the wind blows only horizontally, i.e.
\begin{equation}
\textbf{v} \cdot \textbf{e}_z = 0
\end{equation}
\end{itemize}
Eq.~\ref{Eq:Ass3} now becomes
\begin{equation}
\left(
\begin{array}{c}
\varrho \, ( \partial v_x / \partial t ) \\
\varrho \, ( \partial v_y / \partial t ) \\
0
\end{array}
\right)
=
-
\left(
\begin{array}{c}
\partial p / \partial x \\
\partial p / \partial y \\
\partial p / \partial z
\end{array}
\right)
-
\left(
\begin{array}{c}
0 \\
0 \\
\varrho\,g
\end{array}
\right)
\label{Eq:Ass50}
\end{equation}
That is, with the usual assumptions about geometry we would get the
hydrostatic equation
\begin{equation}
\frac{dp}{dz} = -\varrho\,g
\label{Eq:dpdz}
\end{equation}
{\it without} the usual assumption $\textbf{v}(\textbf{r},t)\equiv
0$. In standard thermodynamics for a macroscopic volume the pressure
$p$ is characterized by one number, not a field. Irreversible
thermodynamics is a (classical) field theory and hydrodynamics is a
special case.
\subsection{Input from thermodynamics}
The equation of state for an ideal gas reads
\begin{equation}
p\,v = {\widetilde R} T
\label{Eq:pvRT}
\end{equation}
where $v$ is the volume of one gram, and ${\widetilde R} = R /
\textrm{(1\,Mol)}$, where $R$ is the usual molar gas constant. Using
the density $\varrho$ we also may write, respectively,
\begin{equation}
\frac{p}{\varrho} = {\widetilde R} T,\ \ \ \
\varrho = \frac{p}{ {\widetilde R} T }
\end{equation}
\subsubsection*{Assumption 6}
\begin{itemize}
\item
The air of the atmosphere obeys an equation of state of an ideal
gas.
\end{itemize}
With $\varrho = p / ({\widetilde R} T)$ inserted into
Eq.~\ref{Eq:dpdz} we have
\begin{equation}
\frac{dp}{dz} =
-
\frac{p\,g}{{\widetilde R} T} =
-
\frac{(\textrm{1\,Mol})\,p\,g}{{\widetilde R} T}
\label{Eq:dpdzpgRT}
\end{equation}
If the molecular mass of the gas is greater, then the decrease of
pressure with increase of height will be greater as well. For a
temperature field $T$ that is constant in space and time this
equation can be integrated.

\subsection{The isothermal atmosphere}
\subsubsection*{Assumption 7a}
\begin{itemize}
\item
We postulate an isothermal atmosphere.
\end{itemize}
Separation of variable gives
\begin{equation}
\frac{d}{dz}
\left( \ln \left( \frac{p}{p_0} \right) \right) =
- \frac{g}{{\widetilde R} T}
\end{equation}
which may be integrated to
\begin{equation}
\left( \ln \left( \frac{p}{p_0} \right) \right) =
- \frac{g}{{\widetilde R} T}
(z - z_0)
\end{equation}
yielding
\begin{equation}
p =
p_0
\exp \left(
- \frac{g}{{\widetilde R} T}
(z - z_0)
\right)
\end{equation}
from which, with $\varrho = p / ( {\widetilde R} T)$, one obtains
the density as a function of height
\begin{equation}
\varrho =
\frac{p_0}{{\widetilde R} T}
\exp \left(
- \frac{g}{{\widetilde R} T}
(z - z_0)
\right)
=
\varrho_0
\exp \left(
- \frac{g}{{\widetilde R} T}
(z - z_0)
\right)
\end{equation}
Thus, with help of these relations and assumptions, we obtain the
barometric height formulas in case of an isothermal atmosphere. The
lapse rates for the pressure and density, respectively, depend on
the molecular mass of the gas, since ${\widetilde R} = R
/(\textrm{1\,Mol})$.
\subsection{The adiabatic atmosphere}
In what follows, we need three relations for the heat differential
form $dQ$, namely
\begin{eqnarray}
dQ &=&
C_v(T)\,dT +
\frac{{\widetilde R}T}{v} dv
\\
dQ &=&
( C_v(T) + {\widetilde R} ) \,dT -
\frac{{\widetilde R}T}{p} dp
\\
dQ &=&
\frac{1} { {\widetilde R} }
\left( C_v \left( \frac{p\,v} { {\widetilde R} } \right) +
{\widetilde R} \right) \, p\,dv +
\frac{1}{{\widetilde R}}\,
C_v
\left( \frac{p\,v}{{\widetilde R}} \right)
v\,dp
\label{Eq:dQ}
\end{eqnarray}
\subsubsection*{Assumption 7b}
\begin{itemize}
\item
For the ideal gas we use the reversible work form $p\,dv$.
\end{itemize}
We now calculate die adiabatic state changes, i.e.\ we set $dQ=0$,
and, separating the variables, we obtain
\begin{eqnarray}
\int_{T_1}^{T_2} \frac{C_v(T)}{{\widetilde R}T}\,dT
&=&
-
\int_{v_1}^{v_2} \frac{dv}{v}
=
\ln \left( \frac{v_1}{v_2} \right),\\
\int_{T_1}^{T_2}
\frac{C_v(T)+{\widetilde R}} {{\widetilde R}T}\,dT
&=&
\int_{p_1}^{p_2} \frac{dp}{p} =
\ln \left( \frac{p_2}{p_1} \right)
\end{eqnarray}
\subsubsection*{Assumption 7c}
\begin{itemize}
\item
The specific heats of an ideal gas are independent of the absolute
temperature.
\end{itemize}
Thus, we can continue our calculations with letters
\begin{eqnarray}
\frac{C_v}{{\widetilde R}} \int_{T_1}^{T_2} \frac{dT}{T}
&=&
\frac{C_v}{{\widetilde R}}
\ln \left( \frac{T_2}{T_1} \right)
=
\ln \left( \frac{v_1}{v_2} \right),
\\
\frac{C_v+{\widetilde R}}{{\widetilde R}} \int_{T_1}^{T_2}
\frac{dT}{T}
&=&
\frac{C_v+{\widetilde R}}{{\widetilde R}}
\ln \left( \frac{T_2}{T_1} \right)
=
\ln \left( \frac{p_2}{p_1} \right),
\end{eqnarray}
For constant heat capacities we integrate the third %
equation (\ref{Eq:dQ}) for $dQ=0$. Since
\begin{equation}
0
=
\frac{1}{{\widetilde R}}
(C_v + {\widetilde R}) p\,dv +
\frac{1}{{\widetilde R}} C_v v\,dp
\end{equation}
is equivalent to
\begin{equation}
0
=
(C_v + {\widetilde R}) p\,dv +
C_v v\,dp
\end{equation}
we get
\begin{equation}
( C_v + {\widetilde R} )
\int_{v_1}^{v_2} \frac{dv}{v} =
 - C_v \int_{p_1}^{p_2} \frac{dp}{p}
\end{equation}
and, therefore,
\begin{equation}
\frac{C_v + {\widetilde R}}{C_v}
\ln \left( \frac{v_2}{v_1} \right) =
\ln \left( \frac{p_1}{p_2} \right)
\end{equation}
With our assumptions, we have
\begin{equation}
C_p - C_v = {\widetilde R}
\end{equation}
Setting
\begin{quote}
\begin{itemize}
\item
$C_p/C_v=\kappa$,
\item
$T_2=T$, $T_1=T_0$,
\item
$p_2=p$, $p_1=p_0$,
\item
$v_2=v$, $v_1=v_0$,
\end{itemize}
\end{quote}
we obtain the well-known %
{\it adiabatic equations of state\/}:
\begin{eqnarray}
\kappa \cdot \ln \left( \frac{v}{v_0} \right)
&=&
\ln \left( \frac{p_0}{p} \right) \\
\left( \frac{v}{v_0} \right) ^\kappa
&=&
\frac{p_0}{p} \\
\frac{p}{p_0}
&=&
\left( \frac{v}{v_0} \right) ^{-\kappa}
\end{eqnarray}
and
\begin{eqnarray}
\frac{C_v}{C_p-C_v} \cdot \ln \left( \frac{T}{T_0} \right)
&=&
\ln \left( \frac{v_0}{v} \right) \\
\frac{1}{\kappa- 1} \ln \left( \frac{T}{T_0} \right)
&=&
\ln \left( \frac{v_0}{v} \right) \\
\left( \frac{T}{T_0} \right) ^{1/(\kappa-1)}
&=&
\frac{v_0}{v}
\end{eqnarray}
and
\begin{eqnarray}
\frac{C_v +{\widetilde R}}{{\widetilde R}} \cdot \ln \left(
\frac{T}{T_0} \right)
&=&
\ln \left( \frac{p}{p_0} \right) \\
\frac{\kappa}{\kappa-1} \cdot \ln \left( \frac{T}{T_0} \right)
&=&
\ln \left( \frac{p}{p_0} \right) \\
\frac{p}{p_0}
&=&
\left( \frac{T}{T_0} \right)^{\kappa/(\kappa-1)}
\end{eqnarray}
These adiabatic equations of state are well-known from standard
textbooks. However, almost never the assumptions are discussed,
under which they hold true.

Now we replace the one-gram volume $v$ by the density $\rho$. %
We get
\begin{eqnarray}
\frac{p}{p_0} =
&=&
\left( \frac{\varrho}{\varrho_0} \right) ^\kappa \\
\frac{\varrho}{\varrho_0}
&=&
\left( \frac{T}{T_0} \right) ^{ 1/(\kappa-1) } \\
\frac{p}{p_0}
&=&
\left( \frac{T}{T_0} \right) ^{ \kappa/(\kappa-1) }
\end{eqnarray}
Rewriting the first of these three equations as
\begin{equation}
p
=
\left( \frac{p_0}{\varrho_0^\kappa} \right) \varrho^\kappa
\label{Eq:B}
\end{equation}
and inserting it into Eq.~\ref{Eq:dpdz}, namely $dp/dz=-\varrho\,g$,
whereby we consider $p$ and $\varrho$ as functions of $z$, we
obtain, applying the chain rule of differential calculus,
\begin{eqnarray}
\frac{dp}{dz}
=
\kappa \frac{p_0}{\varrho_0^\kappa}\varrho^{\kappa-1}
\frac{d\varrho}{dz}
&=&
- \varrho\,g \\
\kappa \frac{p_0}{\varrho_0^\kappa}\varrho^{\kappa-2}
\frac{d\varrho}{dz}
&=&
- g \\
\frac{d}{dz} ( \varrho^{\kappa-1} )
&=&
-
\frac{(\kappa-1)\,g\,\varrho_0^\kappa} {\kappa_0\,p_0}
\end{eqnarray}
Integrating this equation we obtain the density $\varrho$ as a
function of height
\begin{eqnarray}
\varrho^{\kappa-1}
&=&
\varrho_0^{\kappa-1}
-
\frac{ (\kappa-1) \, g \, \varrho_0^\kappa }{ \kappa\,p_0 }
(z-z_0) \\
\varrho(z)
&=&
\varrho_0
\left(
1 - \frac{ (\kappa-1) \, \varrho_0 g }{ \kappa\,p_0 }
(z-z_0)
\right)
^{1/(\kappa-1)}
\end{eqnarray}
From this and Eq.~\ref{Eq:B} we get the pressure $p$ as a function
of height $z$
\begin{equation}
p(z) =
p_0
\left(
1 - \frac{ (\kappa-1) \, \varrho_0 g }{ \kappa\,p_0 }
(z-z_0)
\right)
^{\kappa/(\kappa-1)}
\end{equation}
Inserting this into $p/p_0 = (T/T_0)^{\kappa/(\kappa-1)}$ resp. $T =
T_0 (p/p_0)^{(\kappa-1)/\kappa}$ we obtain the temperature $T$ as a
function of the height $z$
\begin{equation}
T(z) =
T_0
\left(
1 - \frac{ (\kappa-1) \, \varrho_0 g }{ \kappa\,p_0 }
(z-z_0)
\right)
=
T_0
- \frac{ (\kappa-1) \, g }{ \kappa \, {\widetilde R} }
(z-z_0)
\end{equation}
We conclude:
\begin{itemize}
\item
The temperature is decreasing linearly with increasing height.
\end{itemize}
As in the case of an isothermal atmosphere the lapse rate will be
higher if the molecular mass is greater. In the adiabatic case this
slope depends on the adiabatic coefficient $\kappa$. Note, that one
can always weaken the assumption to get similar results (e.g.\ moist
adiabatic approximation and so on).

With our assumptions we may rewrite the constants appearing in the
formula before the height term, i.e.
\begin{eqnarray}
\frac{(\kappa-1)\varrho_0\,g}{\kappa\,p_0}
&=&
\frac{(C_p/C_v-1)\varrho_0\,g}{(C_p/C_v)p_0} \\
&=&
\frac{(C_p-C_v)\varrho_0\,g}{C_pp_0} \\
&=&
\frac{(C_p-C_v)\,g}{C_p{\widetilde R}T_0} \\
&=&
\frac{g}{C_pT_0}
\end{eqnarray}
and
\begin{equation}
\frac{(\kappa-1)\,g}{\kappa\,{\widetilde R}}
=
\frac{(C_p-C_v)\,g}{C_p{\widetilde R}}
=
\frac{g}{C_p}
\end{equation}
Finally, with these coefficients the adiabatic height formulas read
\begin{eqnarray}
\varrho(z)
&=&
\varrho_0
\left(
1 - \frac{g}{C_pT_0}
(z-z_0)
\right)
^{C_v/(C_p-C_v)} \\
p(z)
&=&
p_0
\left(
1 - \frac{g}{C_pT_0}
(z-z_0)
\right)
^{C_p/(C_p-C_v)} \\
T(z)
&=&
T_0 -
\frac{g}{C_p}
(z-z_0)
\end{eqnarray}
%
%
\section{Results}
By combining hydrodynamics, thermodynamics, and imposing the above
listed assumptions for planetary atmospheres one can compute the
temperature profiles of idealized atmospheres. In case of the
adiabatic atmosphere the decrease of the temperature with height is
described by a linear function with slope $-g/C_p$, where $C_p$
depends weakly on the molecular mass. As elucidated in our paper
\cite{GT2007,GT2009} mixtures of gases are covered in the context of
Gibbs thermodynamics. Since the measurable thermo\-dynamic
quantities of a voluminous medium, in particular the specific heat
and the thermodynamic transport coefficients, naturally include the
contribution from radiative interactions, we cannot expect that a
change of concentration of a trace gas has any measurable effect. At
this point, it is important to remember that the barometric formulas
do not hold globally but have only a limited range of validity.
%
%
\section*{Appendix: Relevance to the current climate debate}
\addcontentsline{toc}{section}{Appendix: Relevance to the current
climate debate}
In our falsification paper \cite{GT2007,GT2009} we have shown that
the atmospheric CO$_2$ greenhouse effects \cite{Arrhenius1896} as
taken-for-granted concepts in global climatology do not fit into the
scientific framework of theoretical and applied physics. By showing
that
\begin{itemize}
\item[(a)]
there are no common physical laws between the warming phenomenon in
glass houses and the fictitious atmospheric greenhouse effects
\item[(b)]
there are no calculations to determine an average surface
temperature of a planet
\item[(c)]
the frequently mentioned difference of 33 degrees Celsius is a
meaningless number calculated wrongly
\item[(d)]
the formulas of cavity radiation are used inappropriately
\item[(f)]
the assumption of a radiative balance is unphysical
\item[(e)]
thermal conductivity and friction must not be set to zero
\end{itemize}
the atmospheric CO$_2$ greenhouse effects have been refuted within
the frame of physics \cite{GT2007,GT2009}.

In other words, the greenhouse models are all based on simplistic
pictures of radiative transfer and their obscure relation to
thermodynamics, disregarding the other forms of heat transfer such
as thermal conductivity, convection, latent heat exchange {\it et
cetera\/}. Some of these simplistic descriptions define a \lq\lq
Perpetuum Mobile Of The 2nd Kind\rq\rq\ and are therefore
inadmissable as a physical concept.

In the speculative discussion around the existence of an atmospheric
natural greenhouse effect \cite{NYT} or the existence of an
atmospheric CO$_2$ greenhouse effect it is sometimes stated that the
greenhouse effect could modify the temperature profile of the
Earth's atmosphere. This conjecture is related to another popular
but incorrect idea communicated by some proponents of the global
warming hypothesis, namely the hypothesis that the temperatures of
the Venus are due to a greenhouse effect. For instance, in their
book \lq\lq Der Klimawandel. Diagnose, Prognose, Therapie\rq\rq\
(Climate Change. Diagnosis, Prognosis, Therapy) \lq\lq two leading
international experts\rq\rq, Hans-Joachim Schellnhuber and Stefan
Rahmstorf, present a \lq\lq compact and under\-stand\-able
review\rq\rq\ of \lq\lq climate change\rq\rq\ to the general public
\cite{RahmstorfSchellnhuber2006}. On page 32 they explicitly refer
to the \lq\lq power\rq\rq\ of the \lq\lq greenhouse effect\rq\rq\ on
the Venus.

The claim of Rahmstorf and Schellhuber is that the high venusian
surface temperatures somewhere between 400 and 500 Celsius degrees
are due to an atmospheric CO$_2$ greenhouse effect
\cite{RahmstorfSchellnhuber2006}. Of course, they are not. On the
one hand, since the venusian atmosphere is opaque to visible light,
the central assumption of the greenhouse hypotheses is not obeyed.
On the other hand, if one compares the temperature and pressure
profiles of Venus and Earth, one immediately will see that they are
both very similar. An important difference is the atmospheric
pressure on the ground, which is approximately two orders higher
than on the Earth. At 50 km altitude the venusian atmospheric
pressure corresponds to the normal pressure on the Earth with
temperatures at approximately 37 Celsius degrees. However, things
are extremely complex (volcanic activities, clouds of sulfuric
acid), such that we do not go in details here \cite{venus}.
%
%
\section*{Acknowledgement}
\addcontentsline{toc}{section}{Acknowledgement}
Discussions with Dipl.-Met. Dr. Wolfgang Th\"{u}ne, Dipl-Ing. Paul
Bossert, Gerhard Kramm (University of Alaska, Fairbanks), and Klaus
Ermecke (KE Research) are gratefully acknowledged.
%
%

%
%

\begin{thebibliography}{00}
\addcontentsline{toc}{section}{References}


\bibitem{Berberan1996}
M. Berberan-Santos, E. N. Bodunov, L. Pogliani, %
\lq\lq On the barometric formula\rq\rq, %
Am. J. Phys.
\textbf{65}, %
404 %
(1996)

\bibitem{Riegel1992}
C. A. Riegel (ed. by A. F. C. Bridger), %
\textit{Fundamentals of Atmospheric Dynamics and Thermodynamics} %
(World Scientific, Singapore 1992)


\bibitem{GT2009}
G. Gerlich and R.D. Tscheuschner, %
\lq\lq Falsification Of The Atmospheric CO$_2$ Greenhouse Effects
Within The Frame Of Physics\rq\rq, %
Int. J. Mod. Phys.
\textbf{B23}, %
275 %
(2009)

\bibitem{GT2007}
G. Gerlich and R.D. Tscheuschner, %
\lq\lq Falsification Of The Atmospheric CO$_2$ Greenhouse Effects
Within The Frame Of Physics\rq\rq, %
arXiv:0707.1161

\bibitem{Arrhenius1896}
S. Arrhenius, %
\lq\lq On the Influence of Carbonic Acid in the Air
       Upon the Temperature of the Ground\rq\rq,
\textit{Philosophical Magazine} %
\textbf{41}, %
237 %
(1896)

\bibitem{NYT}
A. C. Revkin, %
\lq\lq Earth Scientists Express Rising Concern Over Warming\rq\rq, %
24 January 2008, %
http://dotearth.blogs.nytimes.com/2008/01/24/earth-scientists-express-rising-concern-over-warming/
{\it (as seen on 25. Januar 2010)\/}

\bibitem{Kramm2009}
G. Kramm, R. Dlugi, M. Zelger, %
\lq\lq Comments on the \lq\lq Proof of the atmospheric greenhouse
effect\rq\rq\ by Arthur P. Smith,\rq\rq %
arXiv:0904.2767

\bibitem{RahmstorfSchellnhuber2006}
S. Rahmstorf and H.J. Schellnhuber, %
\textit{Der Klimawandel - Diagnose, Prognose, Therapie (Climate
Change - Diagnosis, Prognosis, Therapy)} %
(C.H. Beck Wissen, M\"{u}nchen 2008)

\bibitem{venus}
\lq\lq Venus Atmosphere Temperature and Pressure Profiles\rq\rq, 24. Januar 2010, %
http://www.datasync.com/$\widetilde{\phantom{x}}$rsf1/vel/1918vpt.htm


\end{thebibliography}
\end{document}